# Observation of Kondo Effect in a Quadruple Quantum Dots

*Runan Shang[1], Hai-Ou Li[1], Gang Cao[1], Guodong Yu[1], Ming Xiao[1], Tao Tu[1], Guang-Can Guo[1], Hongwen Jiang[2], A.M.Chang[3]\*, and Guo-Ping Guo[1]\*\**

1 Key Laboratory of Quantum Information, University of Science and Technology of China, Chinese Academy of Sciences, Hefei 230026, P.R.China；
Synergetic Innovation Center of Quantum Information & Quantum Physics, University of Science and Technology of China, Hefei, Anhui 230026, P.R.China

2 Department of Physics and Astronomy, University of California at Los Angeles, California 90095, USA.

3 Department of Physics, Duke University, Durham, NC 27708-0305, USA

**ABSTRACT:** We investigate the Kondo effect in a quadruple quantum dot device of coupled-double quantum dots (DQDs), which simultaneously contains intra-DQDs and inter-DQDs coupling. A variety of novel behaviors are observed. The differential conductance dI/dV is measured in the upper DQDs as a function of source drain bias. It is found to exhibit multiple peaks, including a zero-bias peak, where the number of peaks exceeds five. Alternatively, tuning the lower DQDs yielded regions of four peaks. In addition, a Kondo-effect switcher is demonstrated, using the lower DQDs as the controller.

\* Corresponding author: yingshe@phy.duke.edu
\*\* Corresponding author: gpguo@ustc.edu.cn

In the past few years, substantial progress on manipulating heterostructure quantum dot (QD)[1-9] has made it possible to tune system parameters very precisely. This provides a powerful tool for studying the Kondo effect which is an intrinsic many-body phenomenon due to the interplay of localized magnetic impurity and itinerant conduction electrons. Many experimental and theoretical works has been done and provide firm understanding for Kondo effect on single quantum dot[1,16]. On double quantum dots (DQDs) the physics become more diverse. A variety of quantum phenomena are found, such as quantum phase transition[5,6], e.g. the single-triplet transition[4], emergent symmetries[10], etc., which greatly extended our knowledge of the Kondo effect.

In this Letter, we study a quadruple QDs system which consists of two sets of DQDs and three types of interaction: 1. the Kondo coupling between the dots and its corresponding reservoir; 2. the Heisenberg coupling between the two dots within each DQDs; 3. the Ising coupling between the two sets of DQDs. The situations with only Heisenberg or Ising coupling have been studied in DQDs but experimental study of the case with both couplings has never been reported to the best of our knowledge. Capacitively-coupled DQDs (i.e. quadruple QD) systems have been studied by many researchers[11-13], especially in the context of quantum control operation, and they are getting more and more mature. In the following, we report our observation in such a quadruple QD system.

Our sample is fabricated on the $GaAs/Al_{0.3}Ga_{0.7}As$ heterostructure which contains a two-dimensional electron gas (2DEG) lying 100nm below the wafer surface with electron density $2.0 \times 10^{11} cm^{-2}$ and mobility $6 \times 10^4 \ cm^2/V \cdot s$. The sample is placed in a dilution refrigerator which could reach base temperature of 40mK.

Fig1 (d) shows our device structure. Gates 1-5 define a typical series-coupled DQDs structure (upper set of DQDs) by depleting the electrons in 2DEG with Ti/Au top gates. The lower set of DQDs is formed using the gates 6-10. These two sets of DQDs are divided by a pair of horizontal gates 11 and 12 and capacitive coupled to each other. The differential conductance G=dI/dV through the upper and lower set of DQDs is measured using the standard lock-in technique. When the upper and lower DQDs work independently, the negative voltage applied on gates 11 and 12 creates a high barrier for the electrons and the probability for them to tunnel through is negligible. The charge state of a set of DQDs can be represented as (N,M)$_i$ which means there are N electrons in the left dot and M electrons in the right dot and i=U,D distinguishes the upper and lower DQDs. Fig1 (a) and Fig1 (b) illustrate the stability diagram of each DQDs. Fig1 (a) and Fig1 (b) illustrate the stability diagram of each DQDs and the states (N,M)$_U$ and (N,M)$_D$, separately marked as well, is prepared for the operations that control the Kondo effect.

The corresponding parameters can be extracted from the honey-comb diagrams[14]. For the upper DQDs, the charging energies $U_{UL}$ of its left dot and $U_{UR}$ of its right dot have almost the same value of 340GHz. The tunneling rate $\Gamma_{UL}$ from the upper left dot to the reservoir, as estimated from the full width at half maximum (HWFW) of corresponding Coulomb blockade peak, is about 18GHz. For $\Gamma_{UR}$ it is difficult to be determined from obscure Coulomb peak. The coupling between the left and right dot $E_{Um}$ is about 16GHz. For the lower DQDs, $U_{DL}$ is about 590GHz and $U_{DR}$ is about 530GHz. The coupling between the left and right dots of lower set of DQDs is about 15GHz. unfortunately, we are not able to give precise values of $\Gamma_{DR}$ and $\Gamma_{DL}$ also because their respective Coulomb peaks are obscure. And more, we estimate the interactions value between the two sets of DQDs to be 8GHz from the line shift as in Fig1 (c) by sweeping across the charge transition for one double dot while stepping the other through to its charge transition point[11]. Another experimental result is that it seems inevitable that the capacity coupling (16GHz here) and tunneling coupling (7GHz here) exist simultaneously in series-coupled DQDs structure. Indeed, the inter-dot tunneling coupling of the upper DQDs measured by the photon-assisted-tunneling method[15] is about 7GHz, which is half of the value estimated using charging diagram.

Carefully tuning the gates of both the upper and lower DQDs, the most notable feature of the Kondo effect, a zero bias peak (ZBP), is observed in the upper DQDs [bottom curve in Fig2 (c)]. Here the gate voltages are V1-12=-0.63V, -0.7V, -0.65V, -0.7V, -0.945V, -0.81V, -0.4V, -0.55V, -0.4V, -0.84V, -0.8V, -0.8V (V1 stand for voltage on gate1, as so all). Both upper and lower DQDs are formed under this set of gate voltages. Upon increasing the base temperature of the system from 50mK to 800mK, the ZBP height undergoes a monotonous decreasing behavior from the initial highest value $0.082e^2/h$ (background subtracted) as shown in Fig2 (b). This enables us to deduce the Kondo temperature here is between 800-900mK.

To establish that the ZBP is due to the Kondo effect, we change slightly the voltage V3 in the upper DQDs while keeping the other voltages fixed (here V6=-0.774V, V10=-0.885V, other gate voltages are unchanged). As a consequence, the intra-dot coupling in upper set of DQDs (between upper left dot and upper right dot) is modulated as well, which offer us an opportunity to observe the phase transition from atomic Kondo state to the coherent bonding Kondo or the anti-ferromagnetic state[5,6]. Fig2 (a) shows the picture where V3 changes from -0.651V to -0.645V, where a less negative voltage corresponds to a larger intra-dot coupling. We found the result that the splitting between single peak and double peaks occur at about -0.648V. When V3 is adjusted along the positive direction the double splitting peaks become more and more asymmetry. This increase in asymmetry may arise as the electrons wavefunction in upper DQD become asymmetric with the sizable increase in the gate voltage.

For simplicity, we assume the charging energy $U$ of upper DQDs is the average value of $U_{UL}$ and $U_{UR}$ and $\Gamma$ is equals to $\Gamma_{UL}$. Intra-dot coupling t in upper set of DQDs is equaled to $E_{Um} = 16GHz$. Based on our calculation, the type of phase transition we see here can be determined as atomic-Kondo to coherent bonding Kondo[18,19] considering the fact $t/\Gamma$ is close to 1 but $j = 4t^2T_K/U$ is still far away from the critical value of 2.5. The change of intra-dot coupling around the transition is too small to be checked (no matter the capacity or tunneling) for the whole splitting process spans less than 1mV as illustrated by the Fig 2 (a).

It's worth pointing out that while the existence of ZBP has been observed before, on the other hand, the occurrence of at least two additional peaks on each side has not been found in previous studies [indicated by red dashed lines in Fig2 (c) inset]. The temperature dependence is shown in Fig2 (c). A series of dI/dV curves corresponding to different base temperatures (temperature increase from bottom to top) demonstrates the decrease and disappearance with the temperature at 50mK, 60mK, 70mK, 80mK, 100mK, 150mK, 200mK and 250mK.

Existence of the other set of DQDs may result in new phenomena that are beyond what can be observed in a single set of DQDs only. We here focus on the lower set of DQDs: Fig3 (b) provides the details of the charging transition area from $(N+1,M)_D$ to $(N,M+1)_D$ in Fig1 (b) and black dashed lines are used to illustrate the obscure Column peaks. Keeping the upper DQDs voltage (gate1-5=-0.63V, -0.7V, -0.65V, -0.7V, -0.945V), we sweep along the white dashed line from (V6=-0.855, V10=-0.807V)[at $(N,M+1)_D$ state] to (V6=-0.752V, V10=-0.919V)[at $(N+1,M)_D$ state] while measuring the dI/dV of the upper set of DQDs. The result of this measurement is presented in Fig3 (a).

Along the direction of decreasing V10 from -0.807V, an obvious ZBP is observed and the peak height grows up to about $0.08e^2/h$ at V10=-0.85V. At the same time, the other double peaks (indicated by red dashed lines) emerge on the opposite side of zero bias voltage from V10=-0.825V and grow closer when V10 become more negative. Similar experiment features has been observed in semiconductor SET by Kogan et. al.[4] where there is Kondo effect of total spin equals to one. However, our experiment differs from the one by Kogan et. al. and exhibit new phenomena. Firstly, the ZBP still exists and is even enhanced in its height when the double peaks occur, a scenario more similar to what was found by Okazaki et. al.[7], who studied a parallel-coupled DQDs system. Secondly, upon close inspection of the pattern, we can find that the outer double peaks in fact do not evolve along symmetric traces corresponding to the zero bias[17]. Fig3 (c) illustrates the differences using two curves by picking out apex position of respective peak when V10 is in the range [-0.825V,-0.85V]. Note that we have reversed the minus bias to positive value and it can be seen that the gap become larger and larger along the sweeping direction.

Keeping our sweeping process, the splitting of the ZBP occurs as V10 reaches about -0.85V. We speculate this as indicative of a phase transition from a Kondo regime to a bonding-singlet of individual Kondo states, similar to what was observed when tuning V3 in the upper DQDs to change the intra-dot coupling. The height of the leftmost peak is enhanced and become parallel to the middle double peaks while the rightmost peak does not show the same enhancement behavior but it could still be distinguished in Fig3 (a). Neglecting the shift at V10=-0.866V (indicated by black dashed line, we will discuss it later), four peaks (indicated by red dashed lines) appears eventually after the splitting process although the rightmost one is relatively obscure.

To confirm that whether the peaks and splitting arise from the interaction between the upper set and lower set of DQDs or from the capacitive coupling between the upper and lower set of gates, we remove the lower set of DQDs and repeat our experiment. V7 and V9 (double plunge gates of lower set of DQDs) are set to 0 from -0.4V and others are fixed. No quantum dot forms as the current through the source to drain of original lower set of DQDs structure is several hundreds of pA. In setting the lower DQD plunger gates (V7 and V9) to zero, the honey-comb charging diagram of the upper set of DQDs changed minimally. This is seen by directly comparing the diagram with V9=-0.4 and V9=0 (not shown here), where only about a 10mV shift in the upper DQD V5 position, in the negative direction, is found. So we believe that changes on lower plunges gates did not affect the charging diagram of the upper set of DQDs much, which is reasonable since the lower plunge gates are relatively far from the upper set of DQDs and we have only changed the voltage from -0.4V to 0.

We call the two cases I (with lower set of DQDs) and II (without the lower set of DQDs), under the same charging state $(N,M)_U$ of upper set of DQDs, we observed the evolution of the Kondo peaks when V6 and V10 are changed simultaneously as in case I from (V6=-0.73 V10=-0.86V) to (V6=-0.55V V10=-1.04V). The result of dI/dV is drawn into Fig4 a. The ZBP, which spans the whole diagram, is illustrated by a black dashed line. Other two white dashed lines indicate the converging double bumps. There are some obvious differences with case I which are shown in Fig3 (a): Firstly, by removing the lower set of DQDs, there is no splitting nor four peaks feature but only a ZBP which spans the whole V10 range was observed even we change V6 and V10 in a much wider range than in case I. Moreover, the converging peaks collapse from higher peak in case I to lower bump in case II, which suggest the decreasing of Kondo temperature. When V10 is more negative than -0.93V, the converging double peaks vanished. Secondly, comparing the height of ZBP in the range of three peaks existence (ZBP with two converging peaks), we can find that the ZBP reaches $0.086e^2/h$ [see Fig4 (c), background subtracted] in case I while the ZBP is about $0.052e^2/h$ in case II [see Fig4 (b), background subtracted] and keep almost unchanged with V10. The weakening of the Kondo effect can not be explained as coming from the modulation of the lower plunge gates because the peak would be rather higher rather than lower if it is only influenced by the capacitive coupling between the gates. That imply us the ZBP enhancement in case I is due to the existence of lower set of DQDs.

There are many exciting phenomena which have not been observed before, however, we do not aware of any theoretical work that provides a satisfactory explain about the phenomena we report here.

At last we discuss the shift in Fig3 (a) as V6=-0.866V. The charge repopulation between $(N+1,M)_D$ and $(N,M+1)_D$ in the lower set of DQDS will induce a shift of whole diagram of the upper set of DQDS according to the capacitive coupling between the two sets of DQDs. Double curves are extracted from Fig3 (a) at V10=-0.851V and V10=-0.87V are drawn into a new Fig4 (a). We can found that they are almost identical except the most right background. It can be interpreted as that: when we sweeping along a chosen control line in the lower DQDs, a given point in the upper DQDs experiences a corresponding shift in the value of its voltage parameters. When a state transition from $(N+1,M)_D$ to $(N,M+1)_D$ occurs in the control line, the fixed point would dragged back to a former position of the parameters. That provides us a chance to switch Kondo state with a control qubit. In Fig4 (b), we slightly tune the double horizontal peaks from V11=V12=-0.8V to

V11=V12=-0.8003V, no more other physical phenomenon but just a higher ZBP (0.0562$e^2$/h) compare with a rather weaker one (0.0362$e^2$/h) at the opposite contiguous area of the shifting. It is confidential for us to expect the realization of 'on' and 'off' Kondo states switcher if a more refined gate operation is exerted into this quadruple dots system.

In summary, the Kondo effect was observed in an electrically-coupled quadruple dots system. Several unusual features were found: The feature of a ZBP with at least two additional peaks on each side of Vsd has not been previously reported. Moreover, a splitting indicating a transition from an atomic-like Kondo to a coherent single Kondo state appears in the upper DQDs when we tuned the lower DQDs, giving rise to a four peaks feature in the dI/dV. By removing the lower set of DQDs while maintain the gating voltage on the pincher gates controlling intra-dot coupling, we are able to demonstrate that the transition phenomenon arises from inter-dot coupling. To further investigate these novel behaviors will require the application of an in-plane magnetic field, which will be helpful in establishing the physical mechanism responsible for the novel behaviors. Moreover, the intriguing physics due to the competition between the Heisenberg interaction within a set of DQD and the Ising coupling between the two sets of DQD requires more investigations in the future which may greatly improve our understanding of the Kondo effect.

**Acknowledgements:**

This work was supported by the National Fundamental Research Programme (Grant No. 2011CBA00200), and National Natural Science Foundation (Grant Nos. 11222438, 10934006, 11274294, 11074243, 11174267 and 91121014).

**Figure Caption:**

Figure1 (a) The figure shows a honey-comb diagram with four different charge states $(N,M)_U$, $(N+1,M)_U$, $(N,M+1)_U$ and $(N+1,M+1)_U$ (b) The figure shows a honey-comb area with four different charge states $(N, M)_D$, $(N+1, M)_D$, $(N, M+1)_D$ and $(N+1, M+1)_D$ (c) The figure shows the current through upper DQDs by sweeping across the charge transition from $(N,M)_D$ to $(N+1,M)_D$. It is estimated as 8GHz from the shift at V1=-0.785V. (d) The Ti/Au gates 1 to 5 with horizontal gates 11 and 12 define a typical series double quantum dot (upper DQDs) pattern while gate 6-10 (with 11 and 12) define another DQDs (lower DQDs). The black dot 'us' and 'ud' present the source and drain of upper DQDS while 'ds' and 'dd' present the source and drain of lower DQDs. Gates 11 and 12 separate the double DQDs.

Figure2 (a) The figure shows the dI/dV while we change the V3 from -0.651V to -0.645V. A splitting (indicated by white dashed lines) occurs at about V3=-0.647V. (V6=-0.774V, V10=-0.885V, others are seen in text) (b) The curve illustrates the dependence of ZBP with base temperature (the background dI/dV is extracted). We estimate Kondo temperature about 900mK. (c) Successive dI/dV curves (V6=-0.81V, V10=-0.84V, temperature at 50mK, 60mK, 70mK, 80mK, 100mK, 150mK, 200mK and 250mK from bottom curve to top curve). They are off-set by $0.05e^2/h$ and drawn together. The curve is abstract from the white line in inset, which draws dI/dV dependence on V11 (V12 always equals to V11). Five peaks are indicated by five red dashed lines.

Figure3 (a) The figure draws the dI/dV through upper DQDs while sweeping along white dashed line as in (b) and the Kondo peaks are indicated by red dashed line. ZBP with other two converging peaks appear from V10=-0.82V to V10=-0.85V. Then ZBP splits to double peak and a new feature of four Kondo peaks are formed. At V10=-0.866V, there is a shift (indicated by black dashed line) occurs at $\delta = 0$ which is corresponding to the charge number transition of lower DQDs. (b) The figure draws the transition area from $(N+1,M)_D$ to $(N,M+1)_D$. White dashed line indicates the sweeping from (V6=-0.855, V10=-0.807V) to (V6=-0.752V, V10=-0.919V). Black dashed line illustrates the obscure Coulomb peaks and narrower black dashed line indicates the position $\delta = 0$. (c) The figure shows the distance (absolute value) between the left or right peak to ZBP. The increasing gap with V10 from -0.83V to -0.85V illustrates the double peaks are not exactly symmetry.

Fig4 (a) To demonstrate the four peaks feature is due to the existing of lower DQDs (case I), the plunge gates V7 and V9 are set from -0.4V to 0 (case II). Differential conduction dI/dV is read out while sweeping with V6 and V10 simultaneously. The result is drawn into (a) (The shift at V10=-0.875 results from an occasional random shift which should be omitted). The converging double peaks (by white dashed line) has become lower bump but keep the converging behavior from V10=-0.86V to V10=-0.93V as well. They are finally disappeared when V10 is more negative. (b)-(d) are curves extracted separately from [V10=-0.87V in Fig4 (a)] and [V10=-0.835V, V10=-0.895V in Fig3(a)]. In (b), Height of ZBP obviously decreases to $0.052e^2/h$ and keep almost unchanged while whole sweeping process while height of ZBP is $0.086e^2/h$ in (c). There is no splitting appearing in (a). That is entirely different with (d) which illustrates a four peaks feature clearly.

Figure5 (a) Two dI/dV curves extracted from Fig3 (a) with V10=-0.851V and V10=-0.87V are compared, they are almost identical except the right background. (b) Changing the V11 and V12 from -0.8V to -0.8003V, the ZBP experiences a strong step at about V10=-0.866V. A Kondo effect switcher might be realized using the lower set of DQDs as a controller.

Figure 1

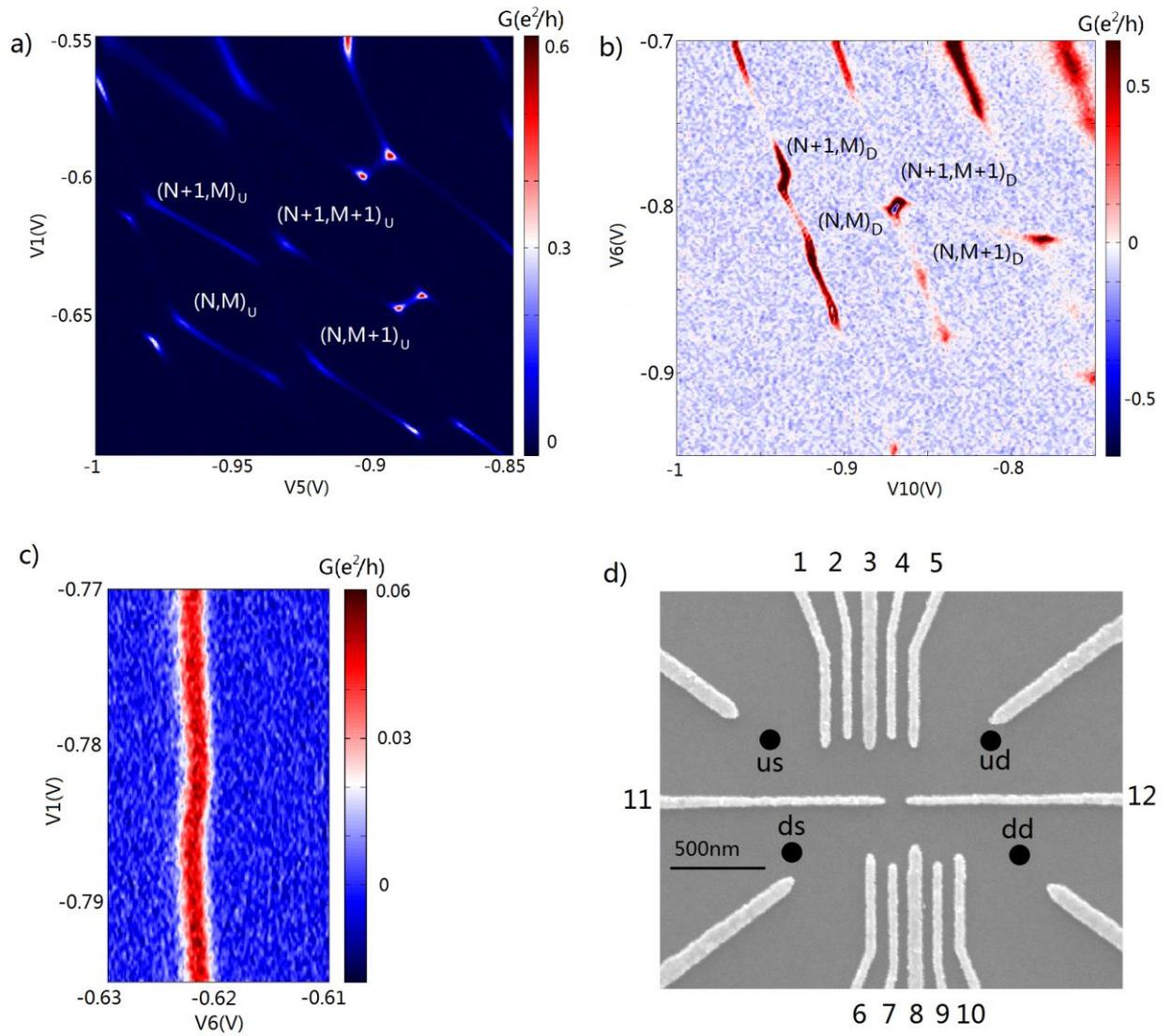

Figure2

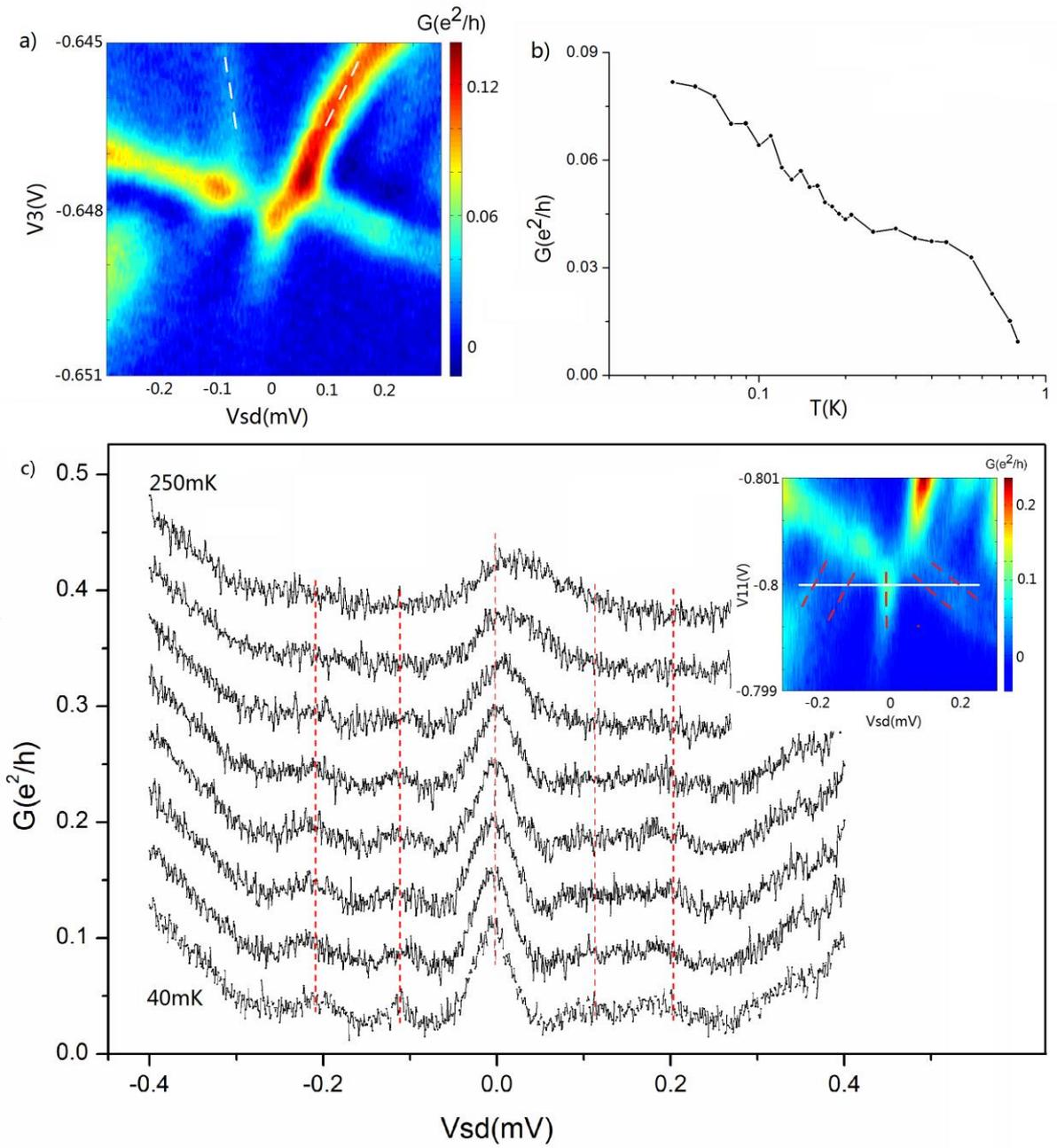

Figure3

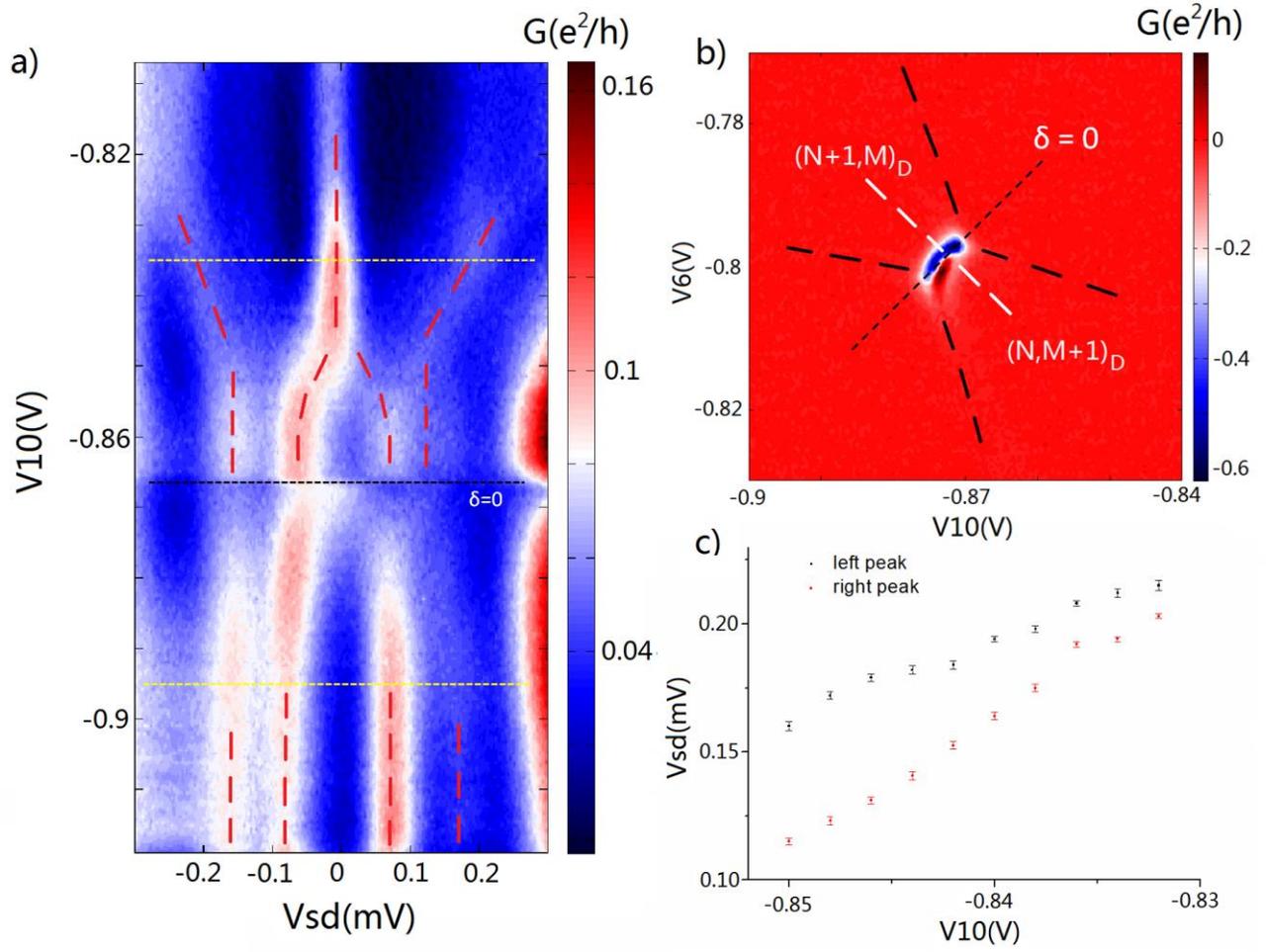

Figure4

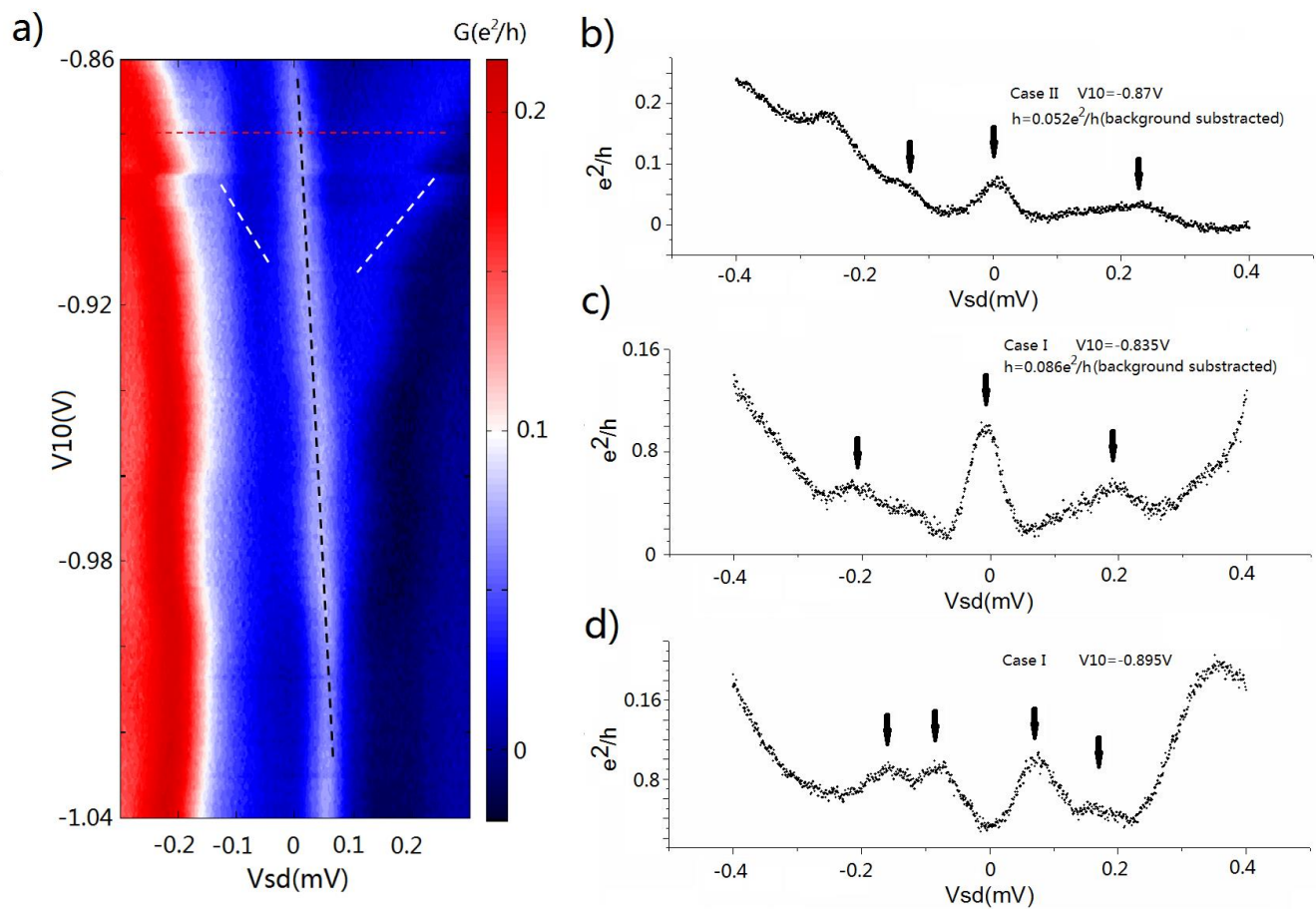

Figure5

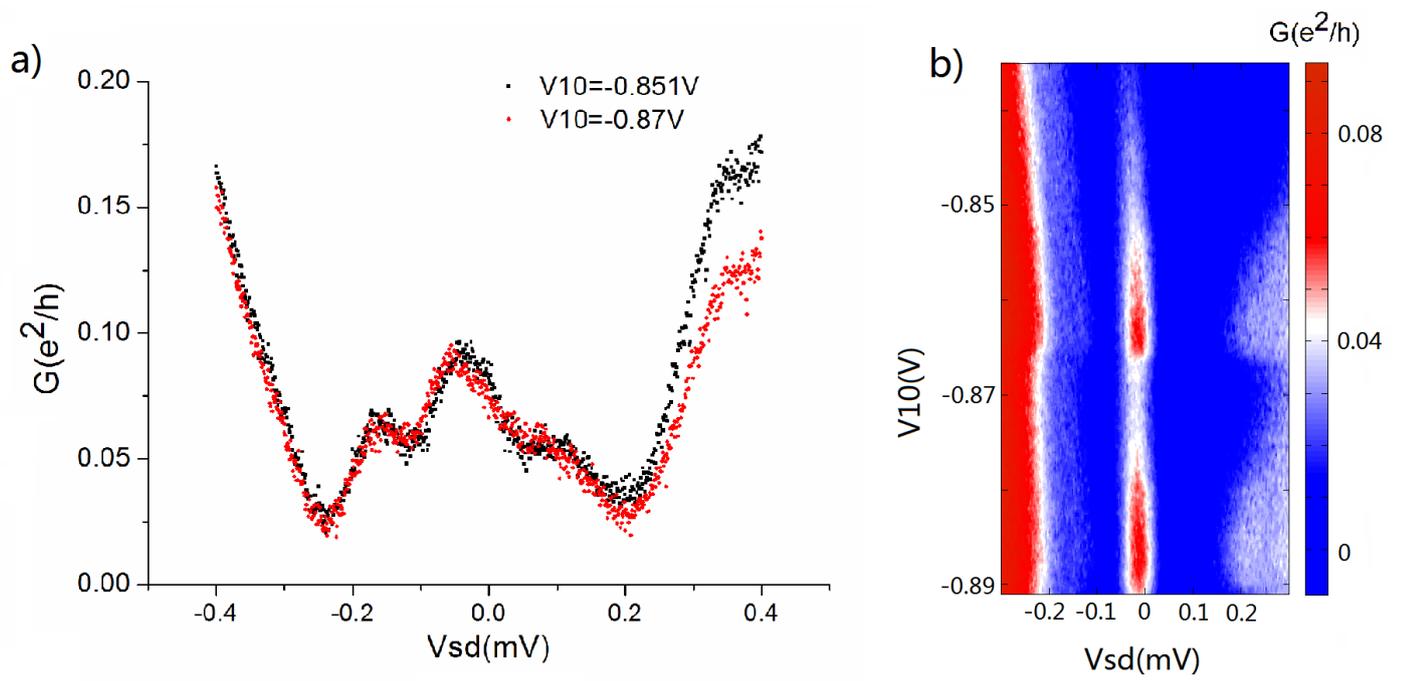